\documentclass{actasen}

\begin{document}

%%ÉèÖÃÊ×Ò³Ò³Âë
\setcounter{page}{1}

\Volume{2014}{1}% Äê¡¢¾í

%%ҳüÉèÖÃ

\runheading{Lu Ji-guang $\&$ Zhou En-ping}%

\title{Two types of glitches in a solid quark star model$^{\dag}$}

\footnotetext{$^{\dag}$ This brief report is based on the article, Zhou E. P., Lu J. G., Tong H. $\&$ Xu R. X., 2014, MNRAS, 443, 2705.

\hspace*{5mm}$^{\bigtriangleup}$ lujig@pku.edu.cn\\

%\noindent 0275-1062/01/\$-see front matter $\copyright$ 2011 Elsevier
%Science B. V. All rights reserved. %%

%\noindent PII:
}

\enauthor{Lu Ji-guang$^{\bigtriangleup}$, Zhou En-ping}{Department of Astronomy, Peking University, Beijing 100871}

\abstract{The glitch of anomalous X-ray pulsars \& soft gamma repeaters (AXP/SGRs) usually
accompanied with detectable energy releases manifesting as X-ray bursts or
outbursts, while the glitch of some pulsars like Vela release negligible energy.
We find that these two types of glitches can naturally correspond to two types of
starquake of solid strange stars. By applying the EoS of quark cluster star and some 
realistic pulsar parameters, we can reproduce consistent results compared with previous constraints and
observations.
}


\maketitle

A glitch is a sudden increase in pulsar¡¯s spin frequency, $\nu$, and the
observed fractions $\Delta\nu/\nu$ range between $10^{-10}$ and $10^{-5}$ (Yu et al. 2013).
In neutron star models, pulsars are thought to be a fluid star with a thin
solid shell. The physical mechanism of glitch is believed to be the coupling
and decoupling between outer crust (rotating slower) and the inner superfluid
(rotating faster) (Anderson $\&$ Itoh 1975; Alpar et al. 1988). However, the absence of evident energy
release during even the largest glitches ($\Delta \nu/\nu \sim 10^{-6}$)
of Vela pulsar is a great challenge to this glitch scenario (G$\mathrm{\ddot{u}}$rkan et al. 2000; Helfand et al. 2001).
At the early years of glitch study, authors believed that glitch mainly happens
on young pulsars, but the glitches detected from AXP/SGRs (anomalous X-ray
pulsars/soft gamma repeaters), usually accompanied with energy release (Kaspi et al. 2003; Tong $\&$ Xu 2011; Dib $\&$ Kaspi 2014).
Thus, the glitch can be divided into two types depending on the energy releasing.
We find that these two types of glitches can naturally correspond to two
types of starquakes of solid stars. The two types of starquakes can be divided
into bulk variable starquake and bulk invariable starquake based on the volume variation.

\begin{figure}[!h]
\centering
{\includegraphics[angle=0,width=12cm]{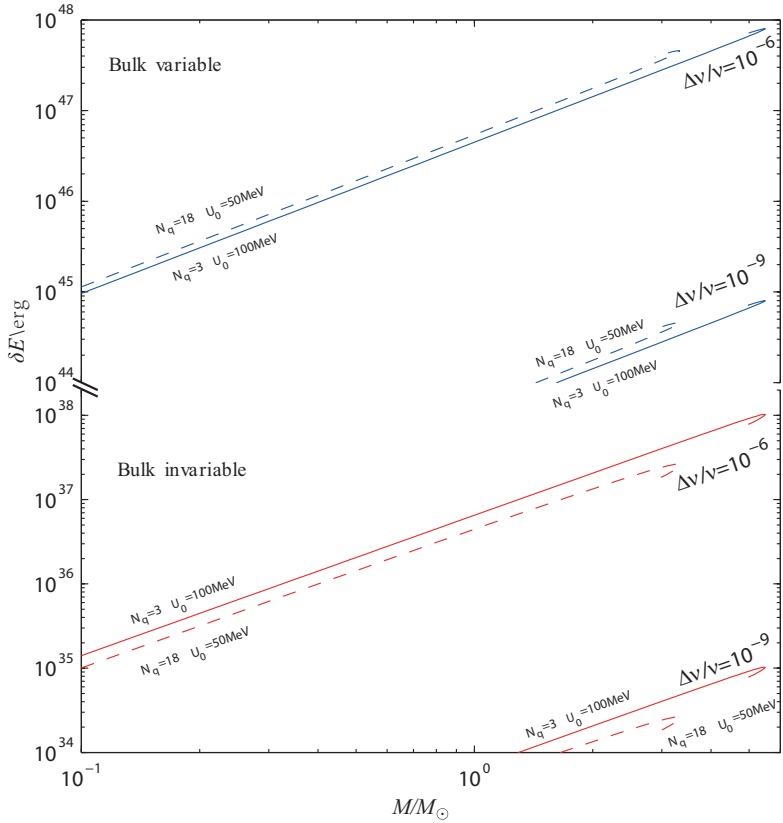}}
%\vspace{-2mm}
\caption{The total energy release during the bulk-variable glitches and bulk-invariable
glitches with amplitudes of $10^{-6}$ and $10^{-9}$.
The Lennard-Jones interaction is applied as an approximation to work out
the mass-radius relation (Lai $\&$ Xu 2009).
There are two main factors in this approximation: the number of quarks in one
cluster ($N_{\mathrm{q}}$) and the depth of the potential($U_{0}$).
The case of 3-quark clusters with potential of 100MeV (solid lines) and 18-quark
clusters with potential of 50MeV (dashed lines) are considered.
It's also worth noting that the energy release during a bulk-invariable
glitch is related to the time intervals between two glitches.
In this calculation the glitch is thought to happen once per month and the
spin down power is calculated according to the observational data of Vela.}
\end{figure}

The bulk variable starquake was induced by the accretion. During the accretion,
the mass of the star increases, and the density-radius relation varies.
If the star is solid, this density-radius relation variation means structure transformation.
But the elasticity of solid body would resist this transformation.
As the accreting carrying on, the elasticity becomes larger, until it gets to its upper limit.
Then the star will collapse.
It leads to a smaller radius and a smaller inertia moment, then the spin frequency will become larger.
With a typical pulsar and glitch parameter, the estimated energy releasing
coincides to the actually observed value of AXP or SGR glitch.
It means that this theory is rational to explain the glitch of AXP/SGRs.

The other type starquake occurs without a bulk variation, and it is induced
by the spin-down of the pulsar.
With a strong magnetic field, the pulsar emits dipole radiation and losses
rotating energy as it spins, and it would spin-down.
If the star is composed with uniform liquid, its shape will be Maclaurin ellipsoid,
and the ellipticity will decrease as the star spin-down.
For a solid star, the shearing force will resist this ellipticity variation.
As the pulsar spin-down, this shearing force becomes larger.
And the star would crush when the shearing force reaching a critical value.
The inertia moment becomes smaller and the rotation speed up.
With some available parameters, the glitch amplitude can be up to $10^{-6}$
even if the epoch between two glitches is about 1 year.

Fig. 1 show our calculating result of the two types of starquakes.
Note that \textbf{there is a break in the vertical scale of about 8 orders of magnitude}.
Actually, for same glitch amplitude, the releasing energy of bulk variable
will be about $\bf{10^{10}}$ times of bulk invariable starquake energy releasing.
Now, we can build the correspondence between the two types of glitches
and starquakes based on the energy release.

In neutron star models, a pulsar is thought to be a fluid star with a thin solid shell.
In fact, so far only quark star and quark cluster star model develop a solid star model.
Then the two types of glitches may be an implication that the pulsar
is composed by quark matter or quark cluster matter.

\acknowledgements{
We would like to thank the pulsar group of PKU for useful discussions.
This work is supported by National Basic Research Program
of China (2012CB821800), National Natural Science Foundation
of China (11225314$\&$11103021) and XTP project XDA04060604}

\end{document}